\documentclass[runningheads,a4paper]{llncs}

\usepackage{amsmath,amssymb}
\usepackage{bm}
\usepackage{graphicx}
\usepackage{ascmac}

\usepackage{url}
\urldef{\mailsa}\path|takano_masanori@cyberagent.co.jp|
\newcommand{\keywords}[1]{\par\addvspace\baselineskip
\noindent\keywordname\enspace\ignorespaces#1}

\begin{document}

\mainmatter

\title{Lightweight Interactions for Reciprocal Cooperation in a Social Network Game}

\titlerunning{ }

\author{Masanori Takano\and Kazuya Wada\and Ichiro Fukuda}
\institute{CyberAgent, Inc., Chiyoda-ku, Tokyo, Japan\\
}

\toctitle{Lightweight Communications for Reciprocal Cooperation in an SNG}
\tocauthor{M. Takano, K. Wada, and I. Fukuda}
\maketitle

\begin{abstract}
The construction of reciprocal relationships requires cooperative interactions during the initial meetings. However, cooperative behavior with strangers is risky because the strangers may be exploiters. In this study, we show that people increase the likelihood of cooperativeness of strangers by using lightweight non-risky interactions in risky situations based on the analysis of a social network game (SNG). They can construct reciprocal relationships in this manner. The interactions involve low-cost signaling because they are not generated at any cost to the senders and recipients. Theoretical studies show that low-cost signals are not guaranteed to be reliable because the low-cost signals from senders can lie at any time. However, people used low-cost signals to construct reciprocal relationships in an SNG, which suggests the existence of mechanisms for generating reliable, low-cost signals in human evolution.

\end{abstract}

\section{Introduction}

Evolutionary game theory research has shown that reciprocal altruism drives the evolution of cooperation~\cite{trivers1971,Lindgren1991,Nowak1993,Nowak2006,Andre2010,Axelrod,Rand2013}.
In this behavior, an individual acts in a manner that temporarily reduces its fitness, while increasing another individual's fitness, with the expectation that the other individual will behave in a similar manner at a later time. 
This behavior has been observed in humans~\cite{Grujic2010,Grujic2012,rand2011} and other primates~\cite{PACKER1977}. 
In addition, the possibility of this behavior has even been suggested in vampire bats~\cite{Wilkinson1990} and fishes~\cite{Bshary2006}. 

Axelrod~\cite{Axelrod} showed that cooperation based on reciprocity requires friendly interactions during the initial meeting in simulations of the iterated Prisoner's Dilemma game.
Because reciprocal cooperators cooperate with individuals who cooperated with them previously.
Indeed, experimental studies using game theory have shown that humans tend to be cooperative in their first meetings without prior interactions~\cite{rand2011,Grujic2012,Wang2012,Peysakhovich2013}.

However, an interaction with strangers can be risky because it is difficult to know each other's levels of cooperativeness.
Therefore, mechanisms for cooperation (kin selection~\cite{Hamilton1963}, direct reciprocity~\cite{trivers1971,Lindgren1991,Nowak1993,Andre2010,Axelrod}, indirect reciprocity~\cite{Nowak2005}, and tags~\cite{Riolo2001}) generate a structured interaction where individuals interact more frequently with acquaintances because strangers may be exploiters.
Nonetheless, humans tend to be cooperative during their first meetings without prior interactions~\cite{rand2011,Grujic2012,Wang2012,Peysakhovich2013}.
The evidences ~\cite{Axelrod,rand2011,Grujic2012,Wang2012,Peysakhovich2013} has been acquired in modeled environments based on the constrained behaviors of humans, or agents, to explicitly analyze their social behavior, e.g., they had to select their strategies without prior interactions.
However, in the real world, we engage in lightweight preliminary interactions, such as observing, eye contact, bowing, and greeting each other.
Therefore, it is important to study these preliminary interactions in a less restrictive environment than that imposed in experimental studies.

In this study, we analyzed the interactions during initial meetings to understand risk reduction behavior in the construction of reciprocal relationships in a social network game (SNG).
In the game, numerous players can behave more freely than possible in the environments used in previous theoretical and experimental studies~\cite{Nowak2006,Rand2013}, i.e., they did not need to select from a sequence of several alternatives because they always had multiple alternatives and the actions of all the players can be recorded. 
In addition, the following features of the SNG make it easier to analyze reciprocal relationships. 
The game allows real players to cooperate and compete with others in situations where the player's benefit is represented by a quantitative value, such as a payoff in game theory.
A previous study~\cite{takano_ngc} demonstrated the existence of reciprocal relationships where cooperators had more advantages than non-cooperators in this SNG.

Many previous studies have used data obtained from interactive online games, particularly in social science~\cite{Castronova2006,Bainbridge2007,Szell2010,Szell2012,Szell2013,Takano2015,takano_ngc,takano_wi_ws}, e.g., the dynamics of virtual world economics~\cite{Castronova2006,Bainbridge2007}, human migration behavior~\cite{Szell2012,Takano2015}, gender differences in social behavior~\cite{Szell2013}, and reciprocal cooperation~\cite{takano_ngc}

\section{Materials and Methods}

In this section, we provide the minimal SNG information and we define cooperative behavior in the SNG (see appendices A.1, A.2, and A.3 for the game information, rules, and definition, respectively).

\begin{figure}
 \begin{center}
  \includegraphics[width=90mm,clip]{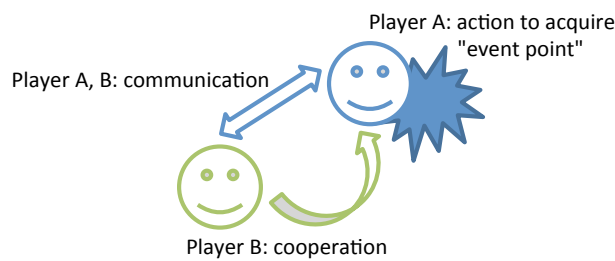}
 \end{center}
 \caption{Overview of players' interaction.
 Player A acts to acquire ``event point'', then player B belonging to a same group can cooperate for A.
 They can communicate each other at any time by using three types of simple text messaging.
 }
 \label{fig_game_abst}
\end{figure}

We analyzed cooperative behavior in the SNG, ``Girl Friend BETA,'' where players acquired ``event points'' and competed in the rankings based on these points because the players received better awards as their rankings increased (Fig. \ref{fig_game_abst}). 
This SNG was released on 10/29/2012.
The player's ranking order was determined by the sum of event points obtained in the period from 3/25/2013 to 4/8/2013.

Players must use their energy to obtain event points; therefore, the number of actions by player is finite. 
There are two methods for replenishing these points: waiting for the points to replenish over time and using a paid item. 
Players must use their resources (items and time) in an effective manner to progress to a higher ranking because their time and money are finite.

Players belong to groups and they must cooperate with each other to play the game efficiently.
The groups are limited to 1--50 players. 
The SNG was designed to ensure that cooperation with group members results in more effective game play. 
We filtered out players who did not belong to groups because most of the active players must belong to groups to play effectively.
Players can create groups on their own. 
Others can apply to join groups at any time and then join a group after the acceptance of their application by an administrator, who was typically a group founder. 
Players can leave a group at any time and apply to join a different group. 
We regarded this behavior as migration.
The migrants were newcomers for the existing group members.
We regarded interactions between migrants (newcomers) and the existing group members within 48 hours of migration as initial meetings.
Players can observe the behavior of members of their group (e.g., attack on common enemies; the details are provided later) because the game system showed their behavior on the game screen.
We targeted groups of five or more active players who logged in at least one or more times to analyze their social interactions. 

Players can communicate at any time using three types of simple text messaging.
The first type is a message from one player to another (direct messaging).
The second type is a message from a player to their group members (group messaging).
The third type is a posting on the forum for their group (forum posting).
These messages have no negative effects on either the senders or receivers, but they also have few or no positive effects \footnote{Players can acquire a few points for a lottery, which provides a card when the players sent messages to each other at the beginning of each day. However, the players had to pay $200$ points to enter the lottery and the effect of the card is small, i.e., the points do not increase the players' abilities.}.
We limited the data to intragroup communication and cooperation.

\begin{table}[hbtp]
 \caption{
 Payoff matrix for the leader game, where $S+T>2R$ and $T>S>R>P$, 
 i.e., Pareto efficiency is achieved when one cooperates and the other does not cooperate. 
 The cooperator then obtains $S$ and the noncooperator receives $T$.}
 \label{table_leader_payoff_matrix}
 \begin{center}
  \begin{tabular}{c||c|c}
    & Cooperation & Noncooperation \\
   \hline \hline
   Cooperation & $R, R$ & $S, T$ \\
   \hline
   Noncooperation & $T, S$ & $P, P$ \\
  \end{tabular}
 \end{center}
\end{table}

We analyzed cooperative behavior in the environment described above. 
It was difficult to track all of the cooperative behaviors because the players can perform various actions in the SNG. 
Thus, we selected a specific cooperative behavior and regarded the frequency of that behavior as a measure of a player's cooperativeness.

We focused on a game scenario where the relationship between players was similar to that in the Leader game (Table \ref{table_leader_payoff_matrix}), but it was not possible for both players to cooperate at the same time in this scenario (see Appendix A.3). 
Pareto efficiency is achieved in the Leader game when one player cooperates and the other do not. 
The cooperator then receives $S$ and the noncooperator receives $T$. 
However, both try to avoid the worst situation (i.e., they receive $P$), but they also do not want to pay the cost for avoiding the worst situation (i.e., they do not want to receive $S$), i.e., the players receive a high payoff by sharing $S$ and $T$ in repeated plays of the game in a process known as $ST$ reciprocity~\cite{Tanimoto2007}. 
We recognized this cooperative behavior, which provided a payoff $T$ from one to the other, as a cooperative behavior in this scenario.

\section{Results}

First, we evaluated the effects of social behavior on the number of cooperation behaviors by others.
We compared the social interactions by migrants (newcomers) within 48 hours of migration and those by existing group members within 48 hours from a random time.
We employed the following generalized linear model (GLM) to analyze these data:
\begin{eqnarray}
\label{reg1}
C'_{i} &\sim& {\rm NB}(\lambda_{i}),  \\
\ln \lambda_{i} &=& \beta_1 \ln \overline{a}_{i} H_{i} + \beta_2 f_{i} + 
             \beta_3 f_i C_i + \beta_4 (1 - f_i) C_i  + \nonumber \\
             && \beta_5 f_i g_i + \beta_6 (1 - f_i) g_i  + 
             \beta_7 f_i G_i + \beta_8 (1 - f_i) G_i  + \nonumber \\
             && \beta_{9} f_i b_i + \beta_{10} (1 - f_i) b_i  + 
              \beta_{11t} d_{ti} + \beta_{12}. \nonumber 
\end{eqnarray}
This model was used to explain the number of cooperative behaviors from group members to player $i$ ($C'_i$) based on a migrant flag $f_{i}$ (if $i$ migrated, then $f_{i}$ = 1; else $f_{i}$ = 0), an interaction between $f_{i}$ and their social behaviors (the number of cooperative behaviors by $i$ ($C_{i}$), the number of direct messages by $i$ ($g_{i}$), the number of group messages by $i$ ($G_{i}$), and the number of forum posts by $i$ ($b_{i}$)), and trends in the cooperative behavior on day $t$ ($d_{ti}$) as dummy variables. 
In addition, we used the log of the product of the number of attacks by the group members $\overline{a}_{i}$ and the number of help requests from $i$ to their group members ($H_{i}$) because this value was expected to increase $C'_{i}$ proportionally if group members cooperated at random (see Appendix A.3), i.e. this controls $i$'s group effect.
$d_{ti}$ was entered as covariates to control for the influence of each day.
${\rm NB}(x)$ shows that $x$ follows a negative binomial distribution.
We estimated its parameters with $80,880$ relationships between players, sampled at random.
We considered this model because the data exhibited over-dispersion when we applied the GLM with a Poisson distribution.

\begin{table}
\caption{
  Results of the regression analysis based on the effects of social behavior relative to the number of cooperative behaviors by others (eq. \ref{reg1}).
  $***$, $**$, and $*$ indicate that the signs of the regression coefficients did not change in Wald-type $99.9\%$, $99\%$, and $95\%$ confidence intervals, respectively (the symbols have the same meaning in the following tables).
  The regression coefficient of $f_i$, $f_i C_i$, $(1-f_i) C_i$, $f_i g_i$, $(1-f_i) g_i$, $f_i G_i$, $(1-f_i) G_i$, and $f_i b_i$ were positive and significant, even after controlling for the other explanatory variables.
  The positive coefficient of $f_i$ shows that newcomers tended to cooperate more than existing group members.
  The regression coefficients of $C_i$, $g_i$, $G_i$, and $b_{i}$ were positive regardless of whether $f_{i}=1$ was or not, and those for $f_{i}=1$ were larger than those for $f_{i}=0$ (excluding the forum posts by existing group members, which was not significant).
}
\label{tbl_cooped_48hour}
\begin{center}
\begin{tabular}{l|rl}
  Explanatory Variable & Regression Coefficient &（Standard Error） \\ \hline \hline
$\ln \overline{a}_i H_i$  & 0.7836054 & $(  0.0052732  )^{ *** }$  \\  \hline
$f_i$ & 4.9450949  & $(  0.0363335 )^{ *** }$  \\  \hline
$f_i C_i$ & 0.1642087 & $(  0.0068994 )^{ *** }$  \\  \hline
$(1-f_i) C_i$ & 0.1018723 & $(  0.0019206 )^{ *** }$  \\  \hline
$f_i g_i$ & 0.0079778 & $(  0.0008424 )^{ *** }$  \\  \hline
$(1-f_i) g_i$ & 0.0004976 & $(  0.0002189 )^{ * }$  \\  \hline
$f_i G_i$ & 0.0941003 & $(  0.0111839 )^{ *** }$  \\  \hline
$(1-f_i) G_i$ & 0.0494162 & $(  0.0069425 )^{ *** }$  \\  \hline
$f_i b_i$ & 0.0170395 & $(  0.0046058 )^{ *** }$  \\  \hline
$(1-f_i) b_i$ & -0.0002771 & $(  0.0015286 )^{  }$  \\  \hline
$d_1$ & 0.3258377 & $(  0.0507144 )^{ *** }$  \\  \hline
$d_2$ & 0.5307506 & $(  0.0508597 )^{ *** }$  \\  \hline
$d_3$ & 0.7443556 & $(  0.0513828 )^{ *** }$  \\  \hline
$d_4$ & 0.7133664 & $(  0.0506845 )^{ *** }$  \\  \hline
$d_5$ & 0.8200644 & $(  0.0500531  )^{ *** }$  \\  \hline
$d_6$ & 0.9167403 & $(  0.0502217 )^{ *** }$  \\  \hline
$d_7$ & 0.9726432 & $(  0.0511331 )^{ *** }$  \\  \hline
$d_8$ & 0.9641601 & $(  0.0520516 )^{ *** }$  \\  \hline
$d_9$ &     1.0840990 & $(  0.0519211 )^{ *** }$  \\  \hline
$d_{10}$  & 0.9394478 & $(  0.0529310 )^{ *** }$  \\  \hline
$d_{11}$  & 0.8924624 & $(  0.0513280 )^{ *** }$  \\  \hline
$d_{12}$  & 1.0503286  & $(  0.0503736 )^{ *** }$  \\  \hline
$d_{13}$  & 1.3767783 & $(  0.0532730 )^{ *** }$  \\  \hline
$d_{13}$  & 2.3644593 & $(  0.0793361 )^{ *** }$  \\  \hline
Intercept & -8.9843031 & $(  0.0666115 )^{ *** }$  \\  
\end{tabular}
\end{center}
\end{table}

Table \ref{tbl_cooped_48hour} shows the results obtained after analyzing the model.
The results demonstrate that reciprocal relationships were constructed between a newcomer and an existing group member, as well as being maintained between existing group members, and that the three types of messages basically supported the reciprocal relationships.
In addition, the cooperative behavior of newcomers and the three types of messages were more important for reciprocal relationships than existing group members.
The results also suggest that sending messages to others may have demonstrated the cooperativeness of players in this SNG, and the construction of reciprocal relationships required more cooperation and communication than the maintenance of reciprocal relationships.

Second, we tested whether the three types of messages showed the cooperativeness of the players.
We analyzed the relationships between the messaging behavior and cooperative behavior of migrants within 48 hours of migration and of the existing group members within 48 hours of a random time.
We employed the following GLM to analyze the results:
\begin{eqnarray}
\label{reg2}
C_{i} &\sim& {\rm NB}(\lambda_{i}), \\
\ln \lambda_{i} &=& \beta_1 \ln a_{i} H'_{i} + \nonumber \\
             && \beta_4 f_i g_i + \beta_5 (1 - f_i) g_i  + 
             \beta_6 f_i G_i + \beta_7 (1 - f_i) G_i  + \nonumber \\
             && \beta_{8} f_i b_i + \beta_{9} (1 - f_i) b_i  + 
             \beta_{10t} d_{ti} + \beta_{11}. \nonumber 
\end{eqnarray}
This model was used to explain the number of cooperative behaviors by player $i$ ($C_i$) based on the interaction between a migrant flag $f_{i}$ (if $i$ migrated, then $f_{i}$ = 1; else $f_{i}$ = 0) and their messaging behavior (the number of direct messages by $i$ ($g_{i}$), the number of group messages by $i$ ($G_{i}$), and the number of forum posts by $i$ ($b_{i}$)), and the trends in cooperative behavior on day $t$ ($d_{ti}$) as dummy variables. 
In addition, we used the log of the product of the number of attacks by player $i$, $a_{i}$, and the number of help requests from their group members ($H'_{i}$) to $i$ because this value was expected to increase $C_{i}$ proportionally if player $i$ cooperated at random (see appendix A.3), i.e. this controls $i$'s group effect.
$d_{ti}$ was entered as covariates to control for the influence of each day.
We estimated its parameters with $80,880$ relationships between players, sampled at random.
${\rm NB}(x)$ shows that $x$ followed a negative binomial distribution.
We employed this model because the data exhibited over-dispersion when we applied the GLM with a Poisson distribution.

\begin{table}
\caption{
  Results of the regression analysis based on the relationships between messaging behavior and cooperative behavior (eq. \ref{reg2}).
  The regression coefficients of $f_i g_i$, $(1-f_i) g_i$, $f_i G_i$, $(1-f_i) G_i$, $f_i b_i$, and $(1-f_i) b_i$ were positive and significant, even after controlling for the other explanatory variables.
  The coefficients of $g_i$, $G_i$, and $b_{i}$ were positive regardless of whether $f_{i}=1$ was or not, and those for $f_{i}=1$ were larger than those for $f_{i}=0$.
}
\label{tbl_coop_48hour}
\begin{center}
\begin{tabular}{l|rl}
  Explanatory Variable & Regression Coefficient &（Standard Error） \\ \hline \hline
$\ln a_i H'_i$  & 0.3797530 & $(  0.0040444  )^{ *** }$  \\  \hline
$f_i g_i$ & 0.1232340 & $(  0.0008722 )^{ *** }$  \\  \hline
$(1-f_i) g_i$ & 0.0109966 & $(  0.0002245 )^{ *** }$  \\  \hline
$f_i G_i$ & 0.3422500 & $(  0.0116456 )^{ *** }$  \\  \hline
$(1-f_i) G_i$ & 0.1180049 & $(  0.0071251 )^{ *** }$  \\  \hline
$f_i b_i$ & 0.2045634 & $(  0.0046320 )^{ *** }$  \\  \hline
$(1-f_i) b_i$ & 0.0474698 & $(  0.0015434 )^{ *** }$  \\  \hline
$d_1$ & 0.1919667 & $(  0.0516471 )^{ *** }$  \\  \hline
$d_2$ & 0.4586802 & $(  0.0516397 )^{ *** }$  \\  \hline
$d_3$ & 0.6671087 & $(  0.0524117 )^{ *** }$  \\  \hline
$d_4$ & 0.6931665 & $(  0.0516168 )^{ *** }$  \\  \hline
$d_5$ & 0.6556934 & $(  0.0512680  )^{ *** }$  \\  \hline
$d_6$ & 0.7007019 & $(  0.0515783 )^{ *** }$  \\  \hline
$d_7$ & 0.6928495 & $(  0.0527448 )^{ *** }$  \\  \hline
$d_8$ & 0.7550984 & $(  0.0534660 )^{ *** }$  \\  \hline
$d_9$ &     0.7368654 & $(  0.0538759 )^{ *** }$  \\  \hline
$d_{10}$  & 0.6794663 & $(  0.0548714 )^{ *** }$  \\  \hline
$d_{11}$  & 0.6930345 & $(  0.0529291 )^{ *** }$  \\  \hline
$d_{12}$  & 0.7449962  & $(  0.0519552 )^{ *** }$  \\  \hline
$d_{13}$  & 0.6131128 & $(  0.0551394 )^{ *** }$  \\  \hline
$d_{13}$  & 0.6985565 & $(  0.0824675 )^{ *** }$  \\  \hline
Intercept & -4.8263750 & $(  0.0550812 )^{ *** }$  \\  
\end{tabular}
\end{center}
\end{table}

Table \ref{tbl_coop_48hour} shows the results obtained after analyzing the model.
The results demonstrate that the messages sent between players basically indicated their cooperativeness.
The results also suggest that the use of messaging by newcomers  indicated greater cooperativeness than that by existing group members.
Thus, the messaging behavior may not have been important for existing group members who had already constructed reciprocal relationships.

\section{Discussion}

In the present study, players constructed reciprocal relationships in a similar manner to those found in studies based on modeled environments~\cite{Axelrod,rand2011,Grujic2012,Wang2012,Peysakhovich2013}.
We showed that lightweight interactions (three types of messages) were important for constructing reciprocal relationships.
The messages involved low-cost signaling because they incurred no costs for the senders and recipients.
Theoretical studies~\cite{Smith1994a,Smith2003} have shown that low-cost signals are not guaranteed to be reliable because the senders can lie at any time using low-cost signals. 
However, we found that the messages sent by players demonstrated their cooperativeness (i.e., their messages were reliable signals) and their messages helped to construct and maintain their reciprocal relationships.
In particular, the messages sent during initial meetings (messages from newcomers to existing group members) were more important than messages between existing group members.
These results suggest that low-cost signals will be reliable in humans.
The signals may be employed to increase the likelihood of cooperativeness by others in risky situations where they are not known to each other.

This evidence for low-cost signaling in humans provides insights into the mechanisms that generate and maintain large societies.
Players probably use low-cost signals as a form of social grooming, which is used to construct and maintain social relationships~\cite{Dunbar2000}. 
Apes, which are closely related to humans, clean each other's fur as a form of social grooming~\cite{Nakamura2003}.
This social grooming incurs high time costs for the groomers and provides hygiene benefits to the recipients of grooming. 
Therefore, their social grooming will work as a reliable signal.
By contrast, social grooming by humans can be low cost such as the three types of messages used in the SNG, as well as gaze grooming~\cite{Kobayashi1997} and one-to-many grooming (e.g., gossip)~\cite{Dunbar}.
The form of social grooming practiced by apes would be too costly for humans because human groups are larger ape groups, so humans must invest time and effort in grooming others in different ways to create social relationships in large groups~\cite{Dunbar2000}. 
Therefore, the evolution of mechanisms that generate reliable signals will have facilitated the evolution of the signature social structures found in humans.

\section*{Acknowledgment}

We are grateful to professor Takaya Arita at Nagoya University, assistant professor Genki Ichinose at Shizuoka University, and master's course Mitsuki Murase at Nagoya University whose comments and suggestions were very valuable throughout this study.

\setcounter{figure}{0}
\setcounter{table}{0}
\renewcommand{\thefigure}{\Alph{figure}}
\renewcommand{\thetable}{\Alph{table}}


\appendix
\section{Appendix}
\subsection{Game Information}

We analyzed cooperative behavior in the SNG, ``Girl Friend BETA.'' Table \ref{table_target_service} presents the game information.
In this SNG, players create individual decks of cards that they collect and then use their decks to perform tasks in the SNG. A powerful deck, constructed from powerful cards, provides an advantage for game play in various situations. The players' primary motivation in the SNG is to obtain powerful cards. Players can obtain powerful cards as top-ranking rewards (see details later) or by casting lots called ``Gacha.''

Players can communicate at any time using three types of simple text messaging.
The first type was a message from one player to another (direct messaging).
The second type was a message from a player to their group members (group messaging).
The third involved posting on the forum for their group (forum posting).
These messages had no negative effects on either the senders or receivers, but they also had few or no positive effects \footnote{Players can acquire a few points for a lottery, which provided a card when the players sent messages to each other at the beginning of each day. However, the players had to pay $200$ points to enter the lottery and the effect of the card was small, i.e., the points did not increase the players' abilities.}.
We limited the data to intragroup communication and cooperation.

\begin{table}
 \caption{Game information}
 \label{table_target_service}
 \begin{center}
  \begin{tabular}{c|c}
   Developer and Publisher & CyberAgent Inc. \\
   \hline
   Service Name & Girl Friend BETA \\
   \hline
   URL & \url{http://vcard.ameba.jp} \\
   \hline
   Event Type & Raid Battle\\
   \hline
   Event Time Period & 3/25/2013 16:00 to 4/8/2013 14:00 \\
   \hline
   Analysis Time Period & 3/25/2013 0:00 to 4/7/2013 23:59 \\
  \end{tabular}
 \end{center}
\end{table}

\subsection{Game Rules}
\begin{figure}
 \begin{center}
  \includegraphics[width=65mm,clip]{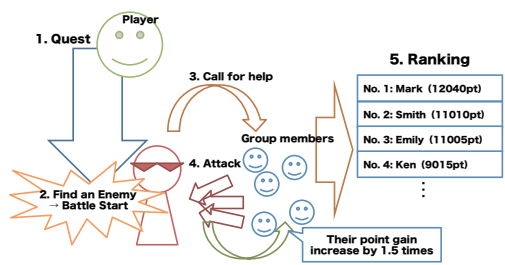}
 \end{center}
 \caption{Overview of raid event. A player conducts ``quests'' to find enemies (1). The player begins a battle upon finding an enemy and then attacks the enemy to obtain points (2). Enemies with very high hit points are strong; thus, they can call for help from other group members whom they have helped to win the battle (3). Players who helped had their point gain increased by $1.5$ times (4). Players compete in rankings based on their points (5).}
 \label{fig_raid}
\end{figure}

Our analysis target was a raid event (Fig. \ref{fig_raid}), in which players attack large enemies\footnote{
The enemy only has hit points as an attribute, meaning that players cannot be attacked by enemies. A player must attack an enemy to acquire event points at the expense of attack points.} and acquire ``event points.'' Players competed in the rankings based on their event points, because they received better awards as their rankings increased.

Players conduct quests\footnote{
This is one of the basic actions in SNGs. 
A player may encounter an enemy on performing certain action.} 
to find enemies during an event. Players begin battles when they find an enemy and then attack the enemy to obtain points. However, enemies with very high hit points are strong, making it difficult for players to win these battles unaided. Thus, they can call for help from other group members, to win the battle. Players who helped had their point gain increased by $1.5$ times. Therefore, players help their fellow group members to acquire more points.

Players' point gains are proportional to the amount of damage caused during attacks, i.e., more powerful decks earn more event points. A player immediately acquires points upon attacking an enemy, even if the enemy is not defeated. However, a player cannot battle another enemy while already battling another enemy, and that enemies' hit points increase with each battle; therefore, players must attack enemies repeatedly in the latter half of an event. Thus, a player who finds an enemy or helps a fellow group member must defeat the enemy before taking a next action, or wait until that the enemy leaves\footnote{
The length of the disable time is set between one and two hours. It is too long to complete the rankings for middle- and higher-rank players, because other players progress in the rankings during their disabled time.
}.

Players increase the amount of damage caused during their attacks by launching ``combo attacks,'' alternate attacks by two or more players in which the players need to launch attacks within ten minutes after other players\footnote{
If a player sequentially attacks an enemy then the attack is not count for the ``combo attacks.'' 
In addition, if players do not attack during ten minutes then their chain of combo attacks are reset to $0$.}.
The longer a chain of combo attacks, the more acquisition points are acquired. Battling enemies together with fellow group members increases the effectiveness of acquisition points.

Players must use a quarter of their attack points to attack; thus, they can attack four times when their point totals are full. There are two methods for replenishing these points: wait for the points to replenish over time or use an item that costs 100 JPY (such items are also sometimes distributed in the game as rewards).

Thus, players must use their resources (items and time) effectively to progress to a higher ranking, e.g., responding to a ``help'' request from their group members to acquire a point gain increase of $1.5$ times, increasing the number of ``combo attacks'' to increase the amount of damage, and reducing the disable time. We defined payment efficiency as the event points per payment, as in game theory.

\subsection{The Test Scenario}

It was impossible to track every cooperative behavior, because players can exhibit various behaviors in the SNG. Hence, we focused on one easily tracked cooperative behavior, and we regarded its frequency as players' cooperativeness.

We focused on the following scenario based on these rules to define players' cooperativeness.
a) An enemy is attacked by a player and fellow group members.
b) The enemy's hit points are very few.
In this scenario, players who defeat the enemy will acquire only a few event points, because their attack power is higher than the enemy's hit points. Thus, their behavior is not efficient for acquiring event points. By contrast, if the players' attack power is lower than the enemy's hit points, their behavior is efficient for acquiring event points. Furthermore, they cannot battle another enemy, if battle with one enemy is ongoing, and therefore must wait until they defeat the enemy to exhibit efficient behavior.

\begin{table}
 \caption{
 Payoff matrix for the test scenario consisting of two players and an enemy with very few hit points. The player who attacks the enemy receives $S$, and the other player receives $T$. If neither player attacks the enemy, then each receives $P$. Attack by both players is impossible, because either player can defeat the enemy.
  }
 \label{table_payoff_matrix}
 \begin{center}
  \begin{tabular}{c||c|c}
    & Attack & Wait \\
   \hline \hline
   Attack & -, - & $S, T$ \\
   \hline
   Wait & $T, S$ & $P, P$ \\
  \end{tabular}
 \end{center}
\end{table}

In simple terms, consider that two players battled an enemy in this scenario, where their relationship is represented in Table \ref{table_payoff_matrix}. 
The relationship between the variables is $T> S > P$ in this payoff matrix. Attack is not efficient, when $S$ is less than $T$. 
However, if they do not attack the enemy, they waste time by waiting for someone else to attack, i.e., $P$ is lowest. 
It is not possible to cooperate both players in this scenario, because an attack on the enemy by either player immediately defeats the enemy. 
The values of this payoff matrix depend on each players situation, e.g., the differences between the two attack powers\footnote{In addition, it does not mean that the relationship between the payoffs is constant. If a player is about to go to sleep, then $S$ is larger than $T$, because the attack points replenish the next morning.}. 
In the scenario, both try to avoid the worst situation (i.e., they get $P$), but they also do not want to pay the cost to avoid the worst situation (i.e., they do not want to get $S$). This social dilemma is similar to the one in the ``Leader game'' (Table \ref{table_leader_payoff_matrix}). In that game, Pareto efficiency is achieved when one cooperates, and the other does not. Then, the cooperator receives $S$, and the noncooperator $T$. 
That is, players receive a high payoff by sharing $S$ and $T$ on repeated plays of the game, a process known as $ST$ reciprocity\cite{Tanimoto2007}. 
We recognized this cooperative behavior, which provided the payoff $T$ from one to the other, as a cooperative behavior in this scenario.

Cooperative behavior is an inefficient attack, as shown in Table \ref{table_payoff_matrix}; thus we define $a_{ij}$ as the attack efficiency indicator: $a_{ij} = e_{ij} / M(\bm{e_{i}})$,
where $e_{ij}$ are the event points in player $i$'s $j$th attack and $M(\bm{e_{i}})$ is the median of $\bm{e_i} = \{e_{i1}, \cdots, e_{iN}\}$ ($N$ is the frequency of player $i$'s attacks). 
We considered cooperative behavior to be in the range of $a \leq 0.40$.
Accordingly, we define $c_i$ as the proportion of cooperative behavior ($a_i \leq 0.40$) for player $i$.
We regarded a cooperator as a player where $c \geq 0.10$.

\end{document}